# Edge-weighted pFISTA-Net for MRI Reconstruction

Jianpeng Cao[1][*]

[1]Department of Electronic Science, School of Electronic Science and Engineering, Xiamen University, Xiamen 361005, China. (*Corresponding author, email: caojianpeng@stu.xmu.edu.cn)

## Abstract

Deep learning based on unrolled algorithm has served as an effective method for accelerated magnetic resonance imaging (MRI). However, many methods ignore the direct use of edge information to assist MRI reconstruction. In this work, we present the edge-weighted pFISTA-Net that directly applies the detected edge map to the soft-thresholding part of pFISTA-Net. The soft-thresholding value of different regions will be adjusted according to the edge map. Experimental results of a public brain dataset show that the proposed yields reconstructions with lower error and better artifact suppression compared with the state-of-the-art deep learning-based methods. The edge-weighted pFISTA-Net also shows robustness for different undersampling masks and edge detection operators. In addition, we extend the edge weighted structure to joint reconstruction and segmentation network and obtain improved reconstruction performance and more accurate segmentation results.

## 1. Introduction

Magnetic resonance imaging (MRI) is the non-invasive imaging modality widely used in clinical diagnosis, but the data acquisition time is relatively long [1]. To shorten imaging time, compressed sensing MRI (CS-MRI) reconstruction methods have been proposed [1-3]. One branch of CS-MRI is to design sparse transform to seek sparse representation of images [1, 4]. However, the iterative process of these methods is relatively time-consuming [5]. Recently, deep learning has become a trend to accelerate MRI reconstruction and can be divided into two categories [5], one based on unrolled algorithms [6-9] and one not [10-14]. Most deep learning methods that are not based on unrolling learn the end-to-end mapping between network inputs and labels [10, 11, 13]. On the other hand, based on prior knowledge and optimized algorithms, the unrolled deep learning approaches unfold the iterative reconstruction process into neural networks [7, 9, 15]. For example, pFISTA-Net [15] unrolls the projected fast iterative soft-thresholding algorithm (pFISTA) [16] into the network that replaces sparse transform with convolutional neural networks (CNN).

The edges of MR images usually contain significant pathological diagnostic information. For example, edges, textures and shapes in brain images can help diagnose Alzheimer's disease [17-19]. To obtain good edge recovery or effectively use edge information, many edge-oriented MRI reconstruction methods have been proposed and can be roughly divided into three categories: multi-modalities reconstruction methods with structural similarity [20-24], edge guided methods for the single modality [25-28], and edge-enhanced deep learning methods [29-31]. (1) Multi-modalities or multi-contrast reconstruction methods utilize similar edge information between different modalities or contrasts as prior knowledge to assist reconstruction. Similar edge structure information includes the location and direction of the edge [20, 23]. Matthias et al. [23] proposed weighted and directional total variation (TV) to integrate structural information from another contrast, such as $T_1$ and $T_2$ images. Three different priors using edge information were proposed for PET-MRI reconstruction, including TV, joint TV and parallel level sets [22]. (2) Edge guided methods determine the weight of each position through the detected edge map, and then carries out weighted TV reconstruction [25-27]. In [27], three different edge detectors were introduced, including Canny [32], local mutual information enhanced detector and wavelet-based detector. Then the influence of different detectors on isotropic and anisotropic weighted TV reconstruction results was discussed. (3) Edge-enhanced deep learning methods learn the end-to-end mapping of MR images and corresponding edge maps during training [29-31]. After the reconstruction output is obtained, the edge detection operator such as sobel operator [33] is employed to detect its edge to obtain the edge map. [29, 30] applied dual discriminator generative adversarial network to simultaneously discriminate the reconstructed output and its corresponding edge map.

Most of the above edge-oriented methods are sensitive to the accuracy of edge detection and those deep learning methods hardly use the edge information directly in the reconstruction process. To tackle this, we propose a novel edge weighted pFISTA-Net. Inspired by the edge weighted TV methods [26, 27], we apply the detected edge map to the soft-thresholding part of pFISTA-Net and adjust the soft-thresholding value of different regions according to the edge map. As for edge detection methods, we discuss three detectors, namely TV based, sobel [33] and Canny operator [32].

The remainder of this paper is organized as follows. In section 2, we detail the network architecture and the designed loss function. Section 3 presents implementation details. Section 4 shows the experimental results. Section 5 provides the discussions on extending to joint reconstruction and segmentation methods and the conclusions are drawn in section 6.

## 2  Methods

pFISTA-Net unrolls pFISTA [16] as the basic structure of the network and specifically, steps in pFISTA are rearranged into the iteration block, which contains two modules: the data consistency (DC) module and the learned operation module. All iteration blocks are sequentially connected to build the network which is like the iterative process of pFISTA. The sampled k-space data is fed into the DC module of each block. The learned operation module consists of three operations: forward operation $P_k$, backward operation $Q_k$ and soft-thresholding operation $T_{\lambda\gamma}$. Different from pFISTA, pFISTA-Net replaces the hand-craft sparse transform with a learnable CNN, namely the forward operation $P_k$. And the backward operation $Q_k$ has the same number of convolution layers as $P_k$. The soft-thresholding operation $T_{\lambda\gamma}$ mimics the soft-thresholding function in pFISTA. Each block has its own soft-thresholding value $\lambda\gamma$ which is learned by the training dataset and the value is equal for all points in the same image, whether it is a smooth region or an edge region. This operation may ignore edge information in the image and further affect the reconstruction quality.

To make better use of edge information, we propose the novel edge weighted pFISTA-Net. As shown in Fig. 1, we can roughly judge whether the region of each pixel in Fig. 1 (a) belongs to the edge from the detected edge maps Fig. 1 (b-d). Inspired by it, we detect edge maps from undersampled images and apply them to the soft-thresholding part of pFISTA-Net. The soft-thresholding value of different regions is adjusted according to the detected edge maps. In this part, we describe the edge weighted pFISTA-Net architecture in detail and its loss function.

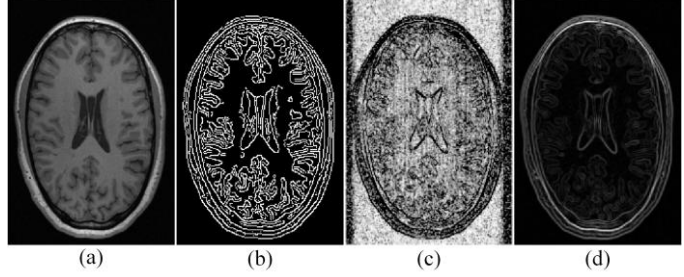

(a)　　　　(b)　　　　(c)　　　　(d)

**Fig. 1.** The fully sampled brain image and the corresponding edge maps. (a) the fully sampled image (b-d) The edge maps obtained by edge detection using Canny, TV and sobel operator, respectively from (a).

**2.1 Network structure**

As Fig. 2 shows, the edge weighted pFISTA-Net mainly includes two modules, namely edge detection module and edge weighted reconstruction module. The detailed description of the two modules are as follows:

1) Edge detection module

As shown in the ED module of Fig. 2 (b), the edge detection module is designed to detect edge maps from undersampled images to obtain edge weight $\mathbf{W}$. Here, we utilize the TV, sobel or Canny operators as the edge detector. To further optimize the obtained edges, we add several convolution layers after edge detection. Finally, the edge weight $\mathbf{W}$ is obtained.

2) Edge weighted reconstruction module

Fig. 2 (c) is the edge weighted reconstruction module which introduces the edge weight $\mathbf{W}$ for soft-thresholding operation. Here, the data consistency part is consistent with the one of pFISTA-Net. $P_m$ and $Q_m$ replace sparse transform and inverse transform with several convolutional layers, respectively. Compared to pFISTA-Net, we modify its soft-thresholding part. $T_{\lambda\gamma\mathbf{W}}$ is the point-wise soft-thresholding operation and can be denoted as

$$T_{\lambda\gamma\cdot\mathbf{W}}(\alpha_i) = \max\left\{|\alpha_i| - \lambda\gamma\frac{1}{w_i + \varepsilon}, 0\right\} \cdot \frac{\alpha_i}{|\alpha_i|}, \tag{1}$$

where $\lambda\gamma$ is the learnable threshold value, $w_i$ is the value of the $i^{th}$ point in $\mathbf{W}$, $\varepsilon$ is a small constant to prevent the denominator from being zero. According to the edge weight $\mathbf{W}$, the points at the edge of the image are given a smaller threshold, and the points of the smooth area are given a larger threshold.

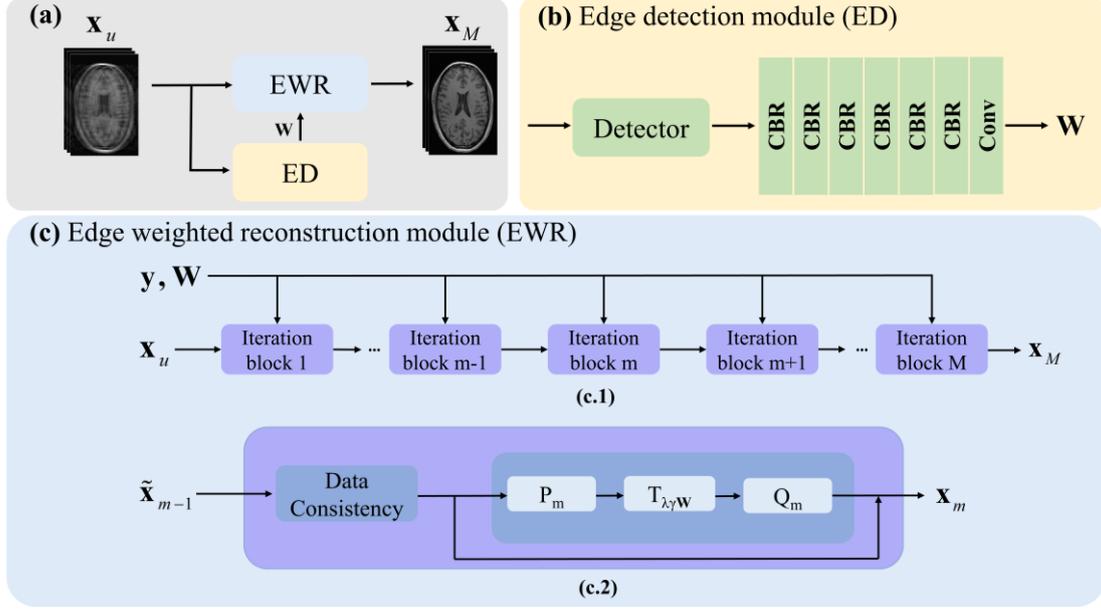

Fig. 2. The overall structure of the proposed edge weighted pFISTA-Net. (a) The flow chart between modules of edge weighted pFISTA-Net, including the edge detection (ED) module and edge weighted reconstruction (EWR) module. (b) The detailed structures of the ED module including edge detection and several convolution layers. (c) The detailed structures of the EWR module. (c.1) is the overview of EWR and (c.2) is the structure of the m$^{th}$ iteration block in (c.1). Note: "CBR" means the convolution, batch normalization, and ReLU. "Conv" means the convolution.

### 2.2 Loss function

Since both EWR and ED modules need to be trained, we adopt the sum of reconstruction loss and edge detection loss as the loss function. For the reconstruction part, we compute the mean squared error between the output of each iteration block $\mathbf{x}_n^m$ in the EWR module and the fully sampled image $\mathbf{x}_n^{ref}$ as follows.

$$\mathcal{L}_{rec} = \sum_{m=1}^{M}\sum_{n=1}^{N} \left\| \mathbf{x}_n^{ref} - \mathbf{x}_n^m \right\|_2^2, \tag{2}$$

where $M$ is the number of iteration blocks in the EWR module, $N$ is the number of training samples and n means the n$^{th}$ training sample, $\mathbf{x}_n^{ref}$ is the label of the n$^{th}$ training sample, $\mathbf{x}_n^m$ is the output of of the m$^{th}$ iteration block of the EWR module.

For the edge detection part, we employ mean squared error between the outputs of the ED module and edge detection results of fully sampled images as follows.

$$\mathcal{L}_{ED} = \sum_{n=1}^{N} \left\| \mathbf{w}_n^{ref} - \mathbf{w}_n \right\|_2^2, \tag{3}$$

where $N$ is the number of training samples, n means the n$^{th}$ training sample, $\mathbf{w}_n^{ref}$ is the edge detection result of the n$^{th}$ fully sampled image and $\mathbf{w}_n$ is the n$^{th}$ outputs of the ED module.

The overall loss is given as

$$\mathcal{L} = \mathcal{L}_{rec} + \mathcal{L}_{ED} = \sum_{m=1}^{M}\sum_{n=1}^{N}\left\|\mathbf{x}_n^{ref} - \mathbf{x}_n^m\right\|_2^2 + \sum_{n=1}^{N}\left\|\mathbf{w}_n^{ref} - \mathbf{w}_n\right\|_2^2, \qquad (4)$$

## 3 Implementation details

### 3.1 Datasets

All experiments were performed on the publicly available Calgary Campinas (CCP) Dataset [34]. It contained fully sampled $T_1$-weighted brain MRI data from 35 subjects acquired on a General Electric (GE) 3T scanner. Data were acquired with a 12-channel imaging coil. The multi-coil k-space data was reconstructed using vendor supplied tools and coil sensitivity maps were normalized to produce a single complex-valued image set that could be back-transformed to regenerate complex k-space samples. We selected a subset of subjects (400 slices in total) for our experiments.

### 3.2 Compared methods and network setup

We compare the proposed method with KIKI-Net [14], pFISTA-Net [15] and HDSLR [35] to demonstrate reconstruction performance. KIKI-Net consists of the data consistency module and cascaded k-space and image domain CNN. Given that the original pFISTA-Net is designed for multi-channel MRI, we implement a single-channel version. HDSLR is an unrolling-based approach that combines the low rank of k-space and image domain priors. All comparison methods were performed according to the typical settings mentioned by the authors.

Considering the training time and reconstruction performance of the proposed method, the block sizes of EWR modules are set to 4. The small constant $\varepsilon$ of the soft-thresholding part is set to 0.1. $\gamma$ and $\lambda$ are initialized as 1 and 0.0001 respectively for each block. All filters are initialized using "Xavier" initialization. Adam is chosen as the optimizer in the training progress and the initial learning rate was set to 0.0001 with exponential decay of 0.99. Considering the memory size, we set the batch size to 4. All experiments were performed on a server equipped with dual Intel Xeon Silver 4210 CPUs, 128 GB RAM, and the Nvidia Tesla T4 GPU (16 GB memory).

### 3.3 Quality evaluation metrics

To quantitatively evaluate the reconstruction performance of the proposed method, we adopt the relative $l_2$ norm error (RLNE) and peak signal-to-noise ratio (PSNR) as the quantitative criteria. The RLNE is calculated by:

$$\mathrm{RLNE} = \frac{\left\|\mathbf{x} - \hat{\mathbf{x}}\right\|_2}{\left\|\mathbf{x}\right\|_2}, \qquad (5)$$

where $\mathbf{x}$ denotes the fully sampled image and $\hat{\mathbf{x}}$ represents the reconstructed image. $\left\|\cdot\right\|_2$ is the $l_2$ norm. The lower RLNE stands for the higher consistency between $\mathbf{x}$ and $\hat{\mathbf{x}}$. The PSNR between $\mathbf{x}$ and $\hat{\mathbf{x}}$ is defined as:

$$\mathrm{PSNR} = 10 \cdot \log_{10}\left(\frac{MN\left\|\mathbf{x}\right\|_\infty}{\left\|\mathbf{x} - \hat{\mathbf{x}}\right\|_2}\right), \qquad (6)$$

where $\mathbf{x}$ and $\hat{\mathbf{x}}$ denote the fully sampled and reconstructed image, respectively. $\left\|\cdot\right\|_2$ is the $l_2$ norm,

$\|\cdot\|_\infty$ is the infinite norm. M and N represent the dimension of the frequency encoding and phase encoding, respectively. The higher PSNR indicates less distortion in the reconstructed image.

## 4. Experimental results

To validate the performance, we design ablation experiments to prove the effectiveness of edge information and compare the proposed with state-of-the-art reconstruction methods. In addition, these methods are compared under different undersampling masks. The proposed methods using TV, sobel and Canny operators to detect edges are named TV-edge, Sobel-edge and Canny-edge, respectively. We further discuss the influence of different detection operators on reconstruction performance.

### 4.1 Ablation study

To verify the effectiveness of introducing edge and edge detection loss, we compare pFISTA-Net, the proposed edge weighted pFISTA-Net with and without edge detection loss. Take the edge detector TV as an example, the proposed method using TV without edge detection loss is named TV-edge-L. As shown in Fig. 3 and Table 1, the proposed methods of introducing edge provide lower error compared with pFISTA-Net. Among them, the reconstruction performance of the method with edge detection loss is slightly better than that of the method without edge detection loss. These results demonstrate the effectiveness of introducing edge and edge detection loss into pFISTA-Net.

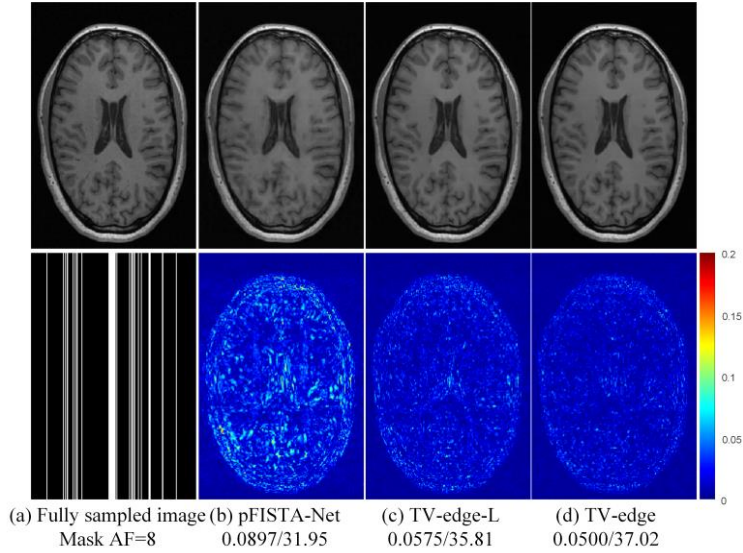

(a) Fully sampled image  (b) pFISTA-Net  (c) TV-edge-L  (d) TV-edge
Mask AF=8                0.0897/31.95    0.0575/35.81   0.0500/37.02

**Fig. 3.** Reconstruction results comparison on 8x accelerated single-channel brain data in the ablation study. (a) the fully sampled image and the Cartesian undersampling pattern with the acceleration factor=8 (AF); (b–d) are the reconstruction results and the corresponding error maps of pFISTA-Net, TV-edge-L and TV-edge, respectively. Note: RLNE/PSNR (dB) are listed below for each reconstruction result.

**Table 1**

RLNE×100/PSNR (dB) of the reconstruction results in the ablation study

| Method | 6-fold | | 8-fold | |
| --- | --- | --- | --- | --- |
| | RLNE×100 | PSNR | RLNE×100 | PSNR |
| pFISTA-Net | 5.70 | 36.07 | 9.31 | 31.84 |
| TV-edge-L | 4.57 | 38.04 | 7.12 | 34.41 |
| TV-edge | **4.47** | **38.23** | **6.51** | **35.27** |

Note: The bold font indicates the lowest mean value of RLNE and the highest mean value of PSNR in the test datasets.

**4.2 Comparison with state-of-the-art methods**

To further verify the performance of the proposed methods, KIKI-Net and HDSLR were selected as comparison methods under two acceleration factors. Reconstructions of the representative images are presented in Fig. 4, in which HDSLR and TV-edge provide images with nice artifacts suppression. However, it can be seen from the error map that the error of TV-edge is lower than that of HDSLR. Table 2 also shows that whether it is six times or eight times undersampling, the reconstruction performance of the method with edges is better than that of other methods.

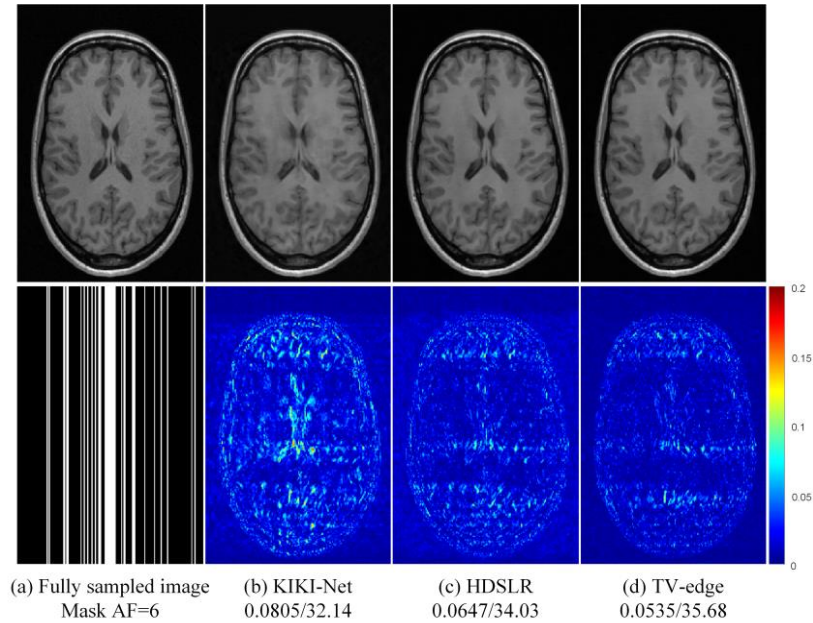

**Fig. 4.** Reconstruction results comparison on 6x accelerated single-channel brain data. (a) the fully sampled image and the Cartesian undersampling pattern with AF=6; (b–d) are the reconstruction results and the corresponding error maps of KIKI-Net, HDSLR and TV-edge, respectively. Note: RLNE/PSNR (dB) are listed below for each reconstruction result.

**Table 2**

RLNE×100/PSNR (dB) of the reconstruction results

| Method | 6-fold | | 8-fold | |
|---|---|---|---|---|
| | RLNE×100 | PSNR | RLNE×100 | PSNR |
| KIKI-Net | 6.01 | 36.13 | 7.74 | 34.89 |
| HDSLR | 5.43 | 36.58 | 8.08 | 33.34 |
| TV-edge | **4.47** | **38.23** | **6.51** | **35.27** |

Note: The bold font indicates the lowest mean value of RLNE and the highest mean value of PSNR in the test datasets.

### 4.3 Study on different undersampling masks

To evaluate the robustness of the proposed methods, we compare the reconstruction metrics of the proposed and the state-of-the-art methods under different undersampling patterns. It can be seen from Fig. 5 and Table 3 that the proposed TV-edge is superior to other methods in artifact suppression and detail preservation, regardless of the undersampling pattern. From the statistical quantitative comparison, the proposed provides lower RLNE and higher PSNR, especially in the case of 2D random undersampling pattern of sampling ratio 18%. Thus, these experimental results indicate that the proposed method has a better reconstruction performance under various undersampling patterns.

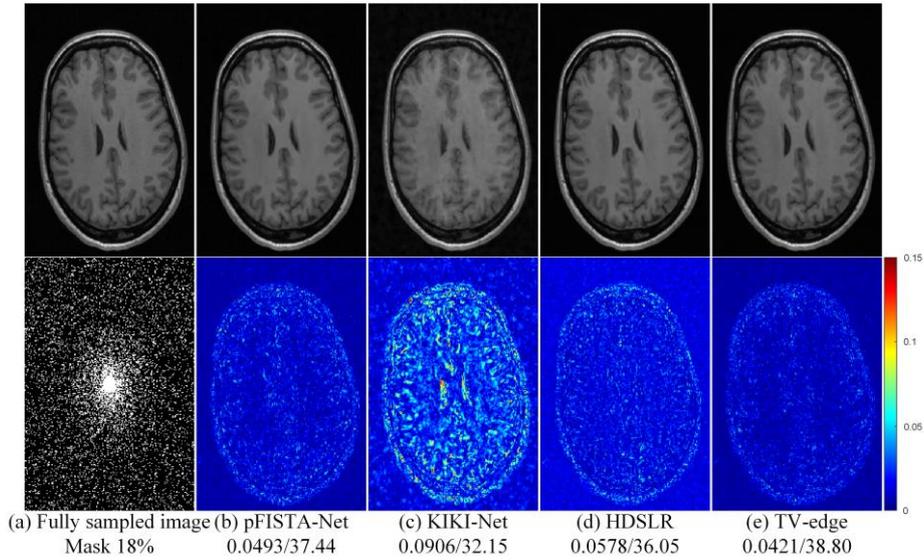

**Fig. 5.** Reconstruction results comparison on single-channel brain data. (a) the fully sampled image and the 2D random undersampling pattern with a sampling rate 18%; (b–e) are the reconstruction results and the corresponding error maps of pFISTA-Net, KIKI-Net, HDSLR and TV-edge, respectively. Note: RLNE/PSNR (dB) are listed below for each reconstruction result.

**Table 3**

RLNE×100/PSNR (dB) of the reconstruction results

| Pattern | Methods | RLNE×100 | PSNR |
|---|---|---|---|
| Random Sampling rate 10% | pFISTA-Net | 6.64 | 34.73 |
| | KIKI-Net | 6.68 | 35.99 |
| | HDSLR | 6.28 | 35.24 |
| | TV-edge | **5.19** | **36.88** |
| Random Sampling rate 18% | pFISTA-Net | 4.89 | 37.38 |
| | KIKI-Net | 6.02 | 36.53 |
| | HDSLR | 5.48 | 36.40 |
| | TV-edge | **4.01** | **39.13** |

Note: The bold font indicates the lowest mean value of RLNE and the highest mean value of PSNR in the test datasets.

### 4.4 Study on different edge detectors

To explore the effect of different edge detection operators on reconstruction performance, we design three detection operators, including TV (TV-edge), Canny operator (Canny-edge) and Sobel operator (Sobel-edge). From Fig. 6 and Table 4, it can be seen that different operators have no obvious impact on the final reconstruction performance and they have improved ability to suppress artifacts. This manifests that the designed network structure is robust to different detection operators.

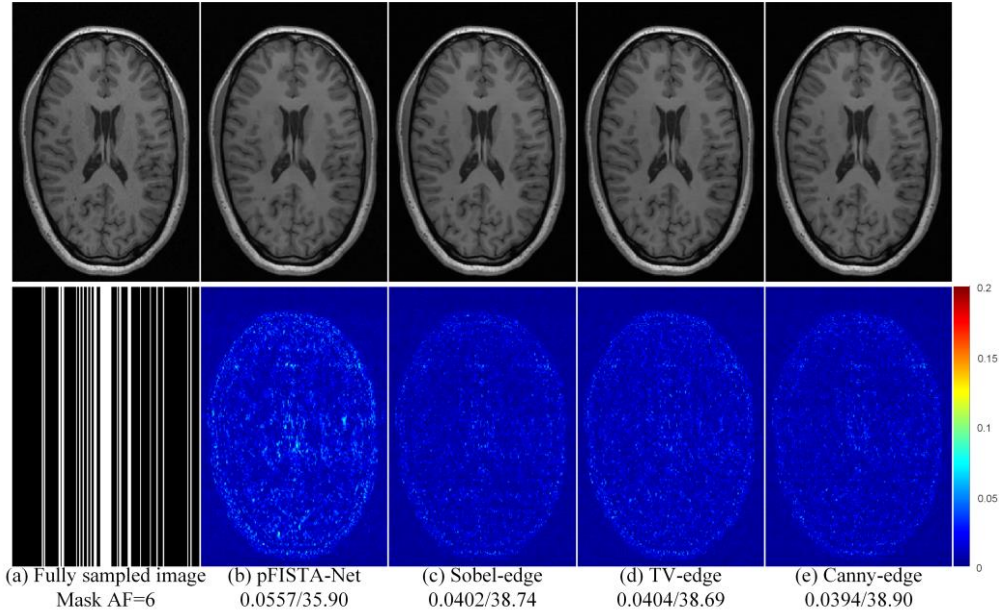

**Fig. 6.** Reconstruction results comparison on 6x accelerated single-channel brain data. (a) the fully sampled image and the Cartesian undersampling pattern with AF=6; (b–e) are the reconstruction results and the corresponding error maps of pFISTA-Net, Sobel-edge, TV-edge and Canny-edge, respectively. Note: RLNE/PSNR (dB) are listed below for each reconstruction result.

**Table 4**
RLNE×100/PSNR (dB) of the reconstruction results

| Method | 6-fold | | 8-fold | |
|---|---|---|---|---|
| | RLNE×100 | PSNR | RLNE×100 | PSNR |
| pFISTA-Net | 5.70 | 36.07 | 9.31 | 31.84 |
| Sobel-edge | 4.50 | 38.20 | 6.93 | 34.68 |
| TV-edge | 4.47 | 38.23 | **6.51** | **35.27** |
| Canny-edge | **4.47** | **38.27** | 6.90 | 34.81 |

Note: The bold font indicates the lowest mean value of RLNE and the highest mean value of PSNR in the test datasets.

# 5. Discussion

After MRI reconstruction, image segmentation can be further discussed [36]. The reconstruction and segmentation methods are usually designed and developed independently, which ignores the potential synergy between the two and consumes more computing power [37]. Inspired by the above proposed method, we design a joint edge-weighted reconstruction and segmentation network (JERS), which applies the edge information obtained from the segmentation results to the soft-thresholding operation of reconstruction. Here we show the network structure, loss function and some experimental results of JERS.

**5.1 Network structure and loss function**

As Fig. 7 shows, JERS mainly includes four modules, namely pre-reconstruction, segmentation, edge detection and edge weighted reconstruction modules. The detailed description of the four modules is as follows:

1) Pre-reconstruction module

As shown in the PR module of Fig. 7 (a), the pre-reconstruction module uses pFISTA-Net to obtain the preliminary reconstruction results $\mathbf{X}_K$, which is to prepare for subsequent segmentation and edge detection and to provide initial input for the edge weighted reconstruction module.

2) Segmentation module

Fig. 7 (b) is the segmentation module. Here, we apply U-Net to brain image segmentation to classify each image pixel into the background, gray matter (GM), white matter (WM), and cerebrospinal fluid (CSF). As shown by $SM_1$ and $SM_2$ in Fig. 7 (a), we perform two segmentations, the first to extract the edges and the second to obtain the final segmentation results.

3) Edge detection module

As shown in the ED module of Fig. 7 (a), the edge detection module is designed to detect obvious edge structures from the segmentation results $\mathbf{S}_1$ to obtain edge weight $\mathbf{W}$. Here, we utilize the sobel operator for edge detection to reduce training costs.

4) Edge weighted reconstruction module

Fig. 7 (c) is the edge weighted reconstruction module identical to the above edge-weighted pFISTA-Net, which introduces edge weight into the soft-thresholding operation of reconstruction.

Finally, we get the reconstruction results $\mathbf{x}_M$ and then go through the segmentation module again to obtain the final segmentation results $\mathbf{S}_2$.

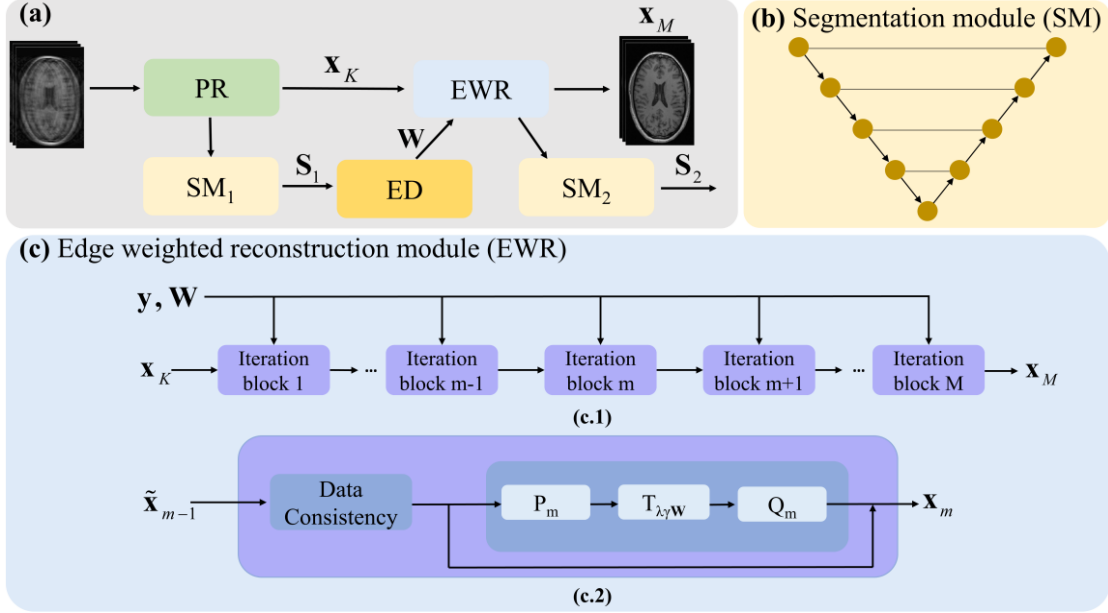

Fig. 7. The overall structure of the proposed JERS. (a) The flow chart between all modules of JERS, including the pre-reconstruction (PR) module, segmentation (SM) module, edge detection (ED) module, and edge weighted reconstruction (EWR) module. (b) The detailed structures of the SM module. (c) The detailed structures of the EWR module. (c.1) is the overview of EWR and (c.2) is the structure of the m$^{th}$ iteration block in (c.1).

In the JERS, we adopt a weighted combination of reconstruction and segmentation losses as the loss function for end-to-end training of JERS. For the reconstruction part, we compute the mean squared error between the output of each iteration block $\mathbf{x}_n^m$ in the PR and EWR module and the fully sampled image $\mathbf{x}_n^{ref}$ as follows.

$$\mathcal{L}_{rec} = \mathcal{L}_{rec}^{PR} + \mathcal{L}_{rec}^{EWR} = \sum_{k=1}^{K}\sum_{n=1}^{N}\left\|\mathbf{x}_n^{ref} - \mathbf{x}_n^k\right\|_2^2 + \sum_{m=1}^{M}\sum_{n=1}^{N}\left\|\mathbf{x}_n^{ref} - \mathbf{x}_n^m\right\|_2^2, \tag{7}$$

where $K$ and $M$ are the number of iteration blocks for PR modules and EWR modules, $N$ is the number of training samples, $\mathbf{x}_n^{ref}$ is the label of the n$^{th}$ training sample, $\mathbf{x}_n^k$ and $\mathbf{x}_n^m$ are the output of the k$^{th}$ iteration block of the PR module and the output of the m$^{th}$ iteration block of the EWR module, respectively.

For the segmentation part, we employ the pixel-wise cross-entropy loss between the outputs of the two SM modules ($\mathbf{S}_1$ and $\mathbf{S}_2$) and the segmentation labels $\mathbf{S}^{ref}$ as follows.

$$\mathcal{L}_{seg} = \mathcal{L}_1 + \mathcal{L}_2 = \sum_{n=1}^{N}\mathbf{S}_n^{ref}\log\mathbf{S}_{1,n} + \sum_{n=1}^{N}\mathbf{S}_n^{ref}\log\mathbf{S}_{2,n}, \tag{8}$$

where $N$ is the number of training samples and n means the $n^{th}$ training sample.

The overall loss is given as

$$\mathcal{L} = \beta_1 \cdot \mathcal{L}_{rec} + \beta_2 \cdot \mathcal{L}_{seg}, \tag{9}$$

where $\beta_1$ and $\beta_2$ are the weight terms regulating the loss function of the reconstruction and segmentation parts.

### 5.2 Reconstruction and segmentation results comparison

Since there was no corresponding segmentation label in the CCP, we chose to generate reference segmentation using FMRIB's Automated Segmentation Tool (FAST) software [38]. It classified the image pixels into GM, WM and CSF through a hidden Markov random field model and an associated Expectation-Maximization algorithm. All the following experiments were conducted on the CCP dataset and the corresponding segmentation labels. It has to be mentioned that although there are several more challenging publicly available segmentation datasets, most of them provide post-processed DICOM images and are not suitable for our task requirements.

As shown in Fig. 8, KIKI-Net (Fig. 8 (b)), HDSLR (Fig. 8 (d)), and the proposed JERS (Fig. 8 (e)) offer promising results while the reconstruction results of pFISTA-Net (Fig. 8 (c)) contain obvious artifacts. According to the error maps, it can be further seen that JERS provides slightly better reconstruction results than KIKI-Net and HDSLR, especially in the brain skull area.

The evaluation criteria performance for two acceleration factors is summarized in Table 5. When the acceleration factor is increased from 6 to 8, the RLNE of all methods is increased and the PSNR is decreased. In both cases, JERS outperforms other state-of-the-art methods, indicating that JERS has the low reconstruction error and good artifact suppression.

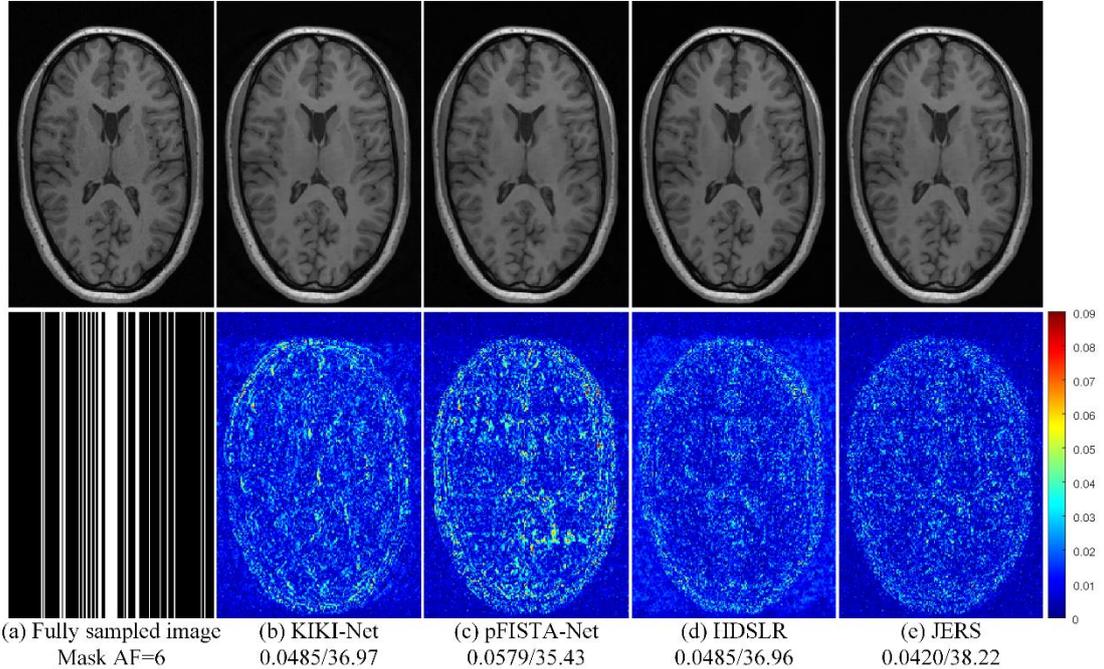

(a) Fully sampled image  (b) KIKI-Net        (c) pFISTA-Net     (d) HDSLR          (e) JERS
    Mask AF=6            0.0485/36.97        0.0579/35.43       0.0485/36.96       0.0420/38.22

**Fig. 8.** Reconstruction results comparison on 6x accelerated single-channel brain data. (a) the fully sampled image and the Cartesian undersampling pattern with AF=6; (b–e) are the reconstruction results and the corresponding error maps of KIKI-Net, pFISTA-Net, HDSLR and JERS, respectively. Note: RLNE/PSNR (dB) are listed for each reconstruction result.

**Table 5**
RLNE×100/PSNR (dB) of the reconstruction results

| Method | 6-fold | | 8-fold | |
|---|---|---|---|---|
| | RLNE×100 | PSNR | RLNE×100 | PSNR |
| KIKI-Net | 6.01 | 36.13 | 7.74 | 34.89 |
| pFISTA-Net | 5.70 | 36.07 | 9.31 | 31.84 |
| HDSLR | 5.43 | 36.58 | 8.08 | 33.34 |
| JERS | **4.86** | **37.66** | **6.92** | **34.96** |

Note: The bold font indicates the lowest mean value of RLNE and the highest mean value of PSNR in the test datasets.

To evaluate the segmentation performance of the proposed JERS, we used the pre-trained U-Net ($U_{FS}$) to test on the reconstruction results of three methods, namely KIKI-Net+$U_{FS}$, pFISTA-Net+$U_{FS}$ and HDSLR+$U_{FS}$. As depicted in Fig. 9 (b) and Table 6, the pre-trained U-Net ($U_{FS}$) provides the most accurate segmentation results among all methods since both the training and test sets of $U_{FS}$ are fully sampled images. In general, the segmentation performance is positively correlated with the quality of the reconstruction results. Probably because of this, the segmentation results obtained by JERS (Fig. 9 (f)) are closer to the segmentation labels compared to KIKI-Net+$U_{FS}$ (Fig. 9 (c)), pFISTA-Net+$U_{FS}$ (Fig. 9 (d)) and HDSLR+$U_{FS}$ (Fig. 9 (e)). From Table 6, it can also be seen that the brain tissues of JERS have higher Dice coefficients.

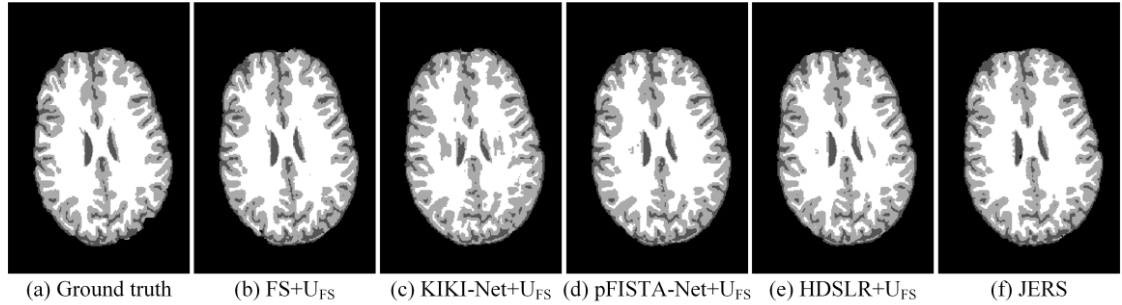

(a) Ground truth    (b) FS+$U_{FS}$    (c) KIKI-Net+$U_{FS}$    (d) pFISTA-Net+$U_{FS}$    (e) HDSLR+$U_{FS}$    (f) JERS

**Fig. 9.** Segmentation results comparison on 6x accelerated single-channel brain data. (a) the ground truth; (b) the segmentation results of $U_{FS}$ tested on fully sampled images; (c–f) the segmentation results of KIKI-Net+$U_{FS}$, pFISTA-Net+$U_{FS}$, HDSLR+$U_{FS}$ and JERS, respectively.

**Table 6**
Dice coefficients for cerebrospinal fluid (CSF), gray matter (GM) and white matter (WM) segmentation results.

| Method | 6-fold | | | 8-fold | | |
|---|---|---|---|---|---|---|
| | CSF | GM | WM | CSF | GM | WM |
| KIKI-Net+$U_{FS}$ | 0.8625 | 0.9016 | 0.9457 | 0.7991 | 0.8566 | 0.9165 |
| pFISTA-Net+$U_{FS}$ | 0.8675 | 0.9036 | 0.9515 | 0.7780 | 0.8085 | 0.8937 |
| HDSLR+$U_{FS}$ | 0.8816 | 0.9196 | 0.9605 | 0.8180 | 0.8517 | 0.9199 |
| JERS | **0.9088** | **0.9522** | **0.9773** | **0.8557** | **0.9073** | **0.9499** |
| | CSF | | GM | | WM | |
| FS+$U_{FS}$ | 0.9329 | | 0.9701 | | 0.9869 | |

Note: The bold font indicates the highest mean value of Dice coefficients in the test datasets other than the $U_{FS}$ results.

## 5.3 Comparison with state-of-the-art method

To verify the performance of the proposed JERS, we compared the state-of-the-art method SegNetMRI [37] and the cascade network containing the pre-reconstruction module that directly cascades pFISTA-Net and U-Net for end-to-end training as shown in Fig. 10 and Table 7. SegNetMRI is the reconstruction and segmentation method and it is built upon a reconstruction network with multiple cascaded blocks each containing an encoder-decoder unit and a data consistency unit, as well as a segmentation network having the same encoder-decoder structure. From the statistical quantitative comparison in Table 7, the proposed JERS outperforms SegNetMRI and the direct cascade method in both reconstruction and segmentation performance, which indicates that JERS can achieve more accurate segmentation while suppressing artifacts.

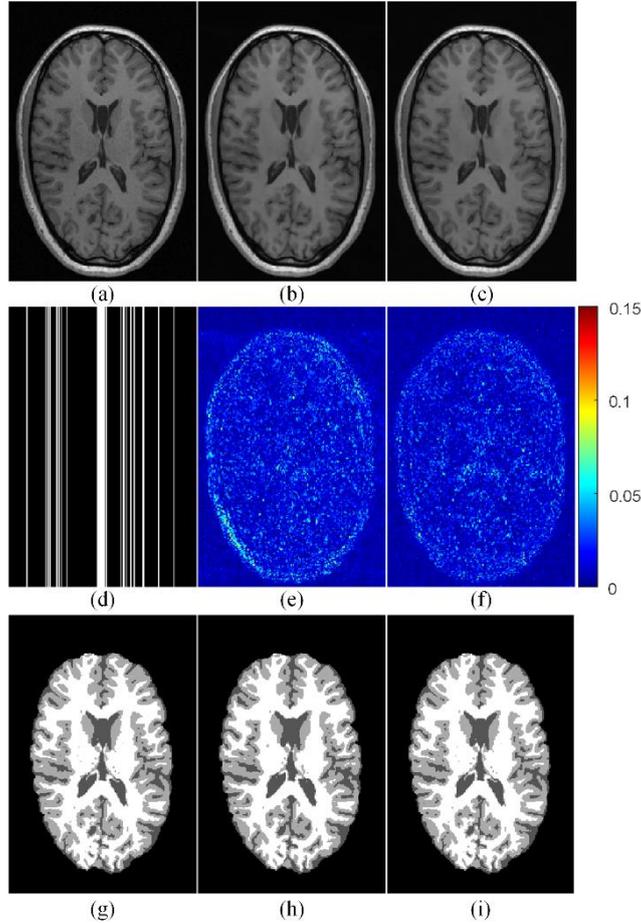

**Fig. 10.** Comparison with SegNetMRI on 8x accelerated single-channel brain data. (a) the fully sampled image; (b–c) are the reconstruction results of SegNetMRI and JERS, respectively; (d) the Cartesian undersampling pattern with AF=8; (e-f) the corresponding error maps of SegNetMRI and JERS, respectively; (g) the segmentation label; (h-j) the segmentation results of SegNetMRI and JERS, respectively.

**Table 7**

Comparison of the reconstruction and segmentation performance of SegNetMRI, the cascaded network and the proposed method JERS. RLNE×100/PSNR (dB) of the reconstruction results and Dice coefficients for CSF, GM and WM segmentation results.

| 6-fold | | | | | |
|---|---|---|---|---|---|
| Method | RLNE×100 | PSNR | CSF | GM | WM |
| Cascade | 8.54 | 32.82 | 0.8253 | 0.8797 | 0.9339 |
| SegNetMRI | 6.67 | 35.86 | 0.8565 | 0.9090 | 0.9531 |
| JERS | **4.86** | **37.66** | **0.9088** | **0.9522** | **0.9773** |
| 8-fold | | | | | |
| Method | RLNE×100 | PSNR | CSF | GM | WM |
| Cascade | 14.26 | 28.24 | 0.5824 | 0.7208 | 0.8461 |
| SegNetMRI | 8.26 | 33.83 | 0.7969 | 0.8581 | 0.9292 |
| JERS | **6.92** | **34.96** | **0.8557** | **0.9073** | **0.9499** |

Note: The bold font indicates the lowest mean value of RLNE, the highest mean value of PSNR and Dice coefficients in the test datasets.

## 6. Conclusion

In this work, we propose a new deep learning method, edge weighted pFISTA-Net, to integrate edge structure information and accelerate magnetic resonance image reconstruction. In the network framework, the edge structure of the image is detected and employed as the soft-thresholding weight of pFISTA-Net. In this way, the soft-thresholding of each point in the reconstruction process can be adjusted according to the region: the points at the edge are given a smaller threshold, and the points of the smooth area are given a larger threshold. We further design the loss function of the sum of reconstruction loss and edge detection loss to continuously improve the accuracy of detected edges. Extensive evaluation experiments demonstrate that the proposed method is effective in adding edge information and edge detection loss, and has higher and more robust reconstruction quality than other methods. Additionally, JERS, which extended the edge weighted structure mentioned above to the joint reconstruction and segmentation network, also shows improved reconstruction performance and more accurate segmentation results.

**Acknowledgment**

The authors would like thank the GPU donated by NVIDIA Corporation. The authors would like to thank Haoming Fang, Chen Qian, Zi Wang, and Xiaobo Qu of Department of Electronic Science of Xiamen University for revising the paper. The authors would like to thank the reviewers for their valuable suggestions on this paper.